# Tipping elements and climate-economic shocks:
# Pathways toward integrated assessment

**Robert E. Kopp[1,2*], Rachael Shwom[2,3], Gernot Wagner[4], Jiacan Yuan[1]**

[1] Department of Earth & Planetary Sciences and Institute of Earth, Ocean & Atmospheric Sciences, Rutgers University, New Brunswick, NJ, USA

[2] Rutgers Energy Institute, Rutgers University, New Brunswick, NJ, USA

[3] Department of Human Ecology, Rutgers University, New Brunswick, NJ, USA

[4] Harvard John A. Paulson School of Engineering and Applied Sciences, Cambridge, MA, USA

Corresponding author: Robert E. Kopp ([robert.kopp@rutgers.edu](mailto:robert.kopp@rutgers.edu))

**Key Points:**

- Potential state shifts in climatic and social systems play an important role in estimates of climate damages.

- 'Tipping points' as broadly understood involve positive feedbacks and abrupt change, but not all climatic thresholds cause abrupt change.

- Some possible climate-economic shocks, whether or not they can be called 'catastrophes', involve tipping points; not all do.

## Abstract

The literature on the costs of climate change often draws a link between climatic 'tipping points' and large economic shocks, frequently called 'catastrophes'. The phrase 'tipping points' in this context can be misleading. In popular and social scientific discourse, 'tipping points' involve abrupt state changes. For some climatic 'tipping points,' the commitment to a state change may occur abruptly, but the change itself may be rate-limited and take centuries or longer to realize. Additionally, the connection between climatic 'tipping points' and economic losses is tenuous, though emerging empirical and process-model-based tools provide pathways for investigating it. We propose terminology to clarify the distinction between 'tipping points' in the popular sense, the critical thresholds exhibited by climatic and social 'tipping elements,' and 'economic shocks'. The last may be associated with tipping elements, gradual climate change, or non-climatic triggers. We illustrate our proposed distinctions by surveying the literature on climatic tipping elements, climatically sensitive social tipping elements, and climate-economic shocks, and we propose a research agenda to advance the integrated assessment of all three.

## 1 Introduction

Whether labeled 'tipping points' [*Lenton and Schellnhuber*, 2007; *Lenton et al.*, 2008], 'large-scale singular events' [*O'Neill et al.*, in rev.; *Smith et al.*, 2001, 2009; *Oppenheimer et al.*, 2014], or 'abrupt impacts' [*National Research Council*, 2002, 2013], large-scale, non-linear shifts in the Earth system are often identified as a reason for concern about climate change [*Oppenheimer et al.*, 2014] and a potential trigger of major economic losses, often described as 'economic catastrophes' [*Nordhaus and Boyer*, 2000; *Wagner and Weitzman*, 2015]. Yet – as the multiple, nearly synonymous phrases suggest – the language used to describe such shifts is unsettled, and the link between physical changes and their socio-economic consequences is often unclear. To enhance cross-disciplinary communications between natural and social scientists working to assess the risks posed by climate-related tipping points and associated costs, this paper seeks to clarify terminology, critically review the underlying literature, and identify research directions in the integrated assessment of these large-scale changes.

The phrase 'tipping point' appears to have originated in industry, where it referred literally to the point at which an engineered system, such as a rail wagon of coal in a Yorkshire foundry [*Burnley*, 1871] or a cup in a tilting water meter [*Hoadley*, 1883], tipped over and emptied its contents. The earliest use of the term in academic research occurred in the social sciences. In a *Scientific American* article, the political scientist Morton Grodzins [1957] applied the phrase 'tip point' to a critical proportion of non-whites in a neighborhood, above which the fraction of whites precipitously declines to zero. 'Tip point' and especially its variant 'tipping point' spread rapidly through the academic and policy literature on neighborhood segregation. (See the growth in the frequency of the phrase in the 1960s and 1970s shown in Figure 1a). *Schelling* [1971] introduced mathematical models of stability points in neighborhood segregation and discussed the conditions that could lead to the development of tipping points.

Malcolm Gladwell popularized the term 'tipping point' in a *New Yorker* article [1996] and in his book *The Tipping Point* [2000], which identified several other examples, including the sudden increase in popularity of Hush Puppies shoes in 1995 and the sharp decline of crime in New York City in the mid-1990s. *Gladwell* [2000] summarized the characteristics of tipping points as (1) being contagious and (2) involving a large change that (a) results from small changes and (b) occurs quickly. (Note the exponential growth of the frequency of 'tipping point' between 1997 and 2007, with a doubling time of ~2.4 years; Figure 1). The term's emergence in climate research followed the Gladwell-inspired trend: *Russill and Nyssa* [2009] note a tipping point in the use of the phrase 'tipping point' in both popular and technical communications about climate change, reached around 2005-2007 amid the broader trend seen in Figure 1. (See the growth of academic publications using the phrases 'climate change' and 'tipping points' in Figure 1b.)

The examples Gladwell provided of tipping points – epidemics of disease, crime, consumption, or behavior – all exhibit rapid shifts between states: specifically, from a state in which an infection or behavior is rare to one in which it is widespread. The scale and speed of the shifts result from positive feedbacks, in particular those related to network effects: the abundance of a contagious element increases the rate at which it spreads, which further increases its abundance. In ecology, such shifts are also known as 'ecological thresholds' or 'regime shifts' [e.g., *Andersen et al.*, 2009], where "ecological regime shifts can be defined as abrupt changes … leading to rapid ecosystem reconfiguration between alternative states" [*Andersen et al.*, 2009, p. 49]. As in the social sciences, a rapid state shift is a key part of the ecological definition.

*Lenton et al.* [2008] formalized the concept of 'tipping points' in the climate system in a way that loosened this definition. *Lenton et al.* [2008] defined a 'tipping element' as a subsystem of the Earth system, subcontinental or larger, that small perturbations can shift into multiple different stable states. A tipping element's tipping point is a critical threshold at which "a small change in forcing triggers a strongly nonlinear response in the internal dynamics of part of the climate system, qualitatively changing its future state" [*Lenton*, 2011, p.201]. *Lenton* [2013] noted that the triggering forcing might arise as a result of the level of forcing, the rate of forcing, or system noise.

Figure 2 illustrates the concept of a tipping element. In response to the changing forcing shown in Figure 2a, a tipping element (Figure 2e-f) changes linearly with forcing until the forcing reaches a critical threshold (indicated by the dotted vertical line). Once the forcing crosses the critical threshold ('tipping point'), the tipping element undergoes a strongly nonlinear state shift. As shown in Figure 2g-h, tipping elements often exhibit hysteresis, so the forcing that triggers a transition from the first state to the second state may differ from the forcing that triggers a transition back into the first state.

In the literal example of the rail wagon of coal, the wagon itself would be the tipping element; the point at which the wagon's physical dynamics commit it to falling on its side and emptying its contents would be the tipping point. Foundry workers might provide the intended forcing to tip the wagon; on poorly constructed tracks, system noise might cause it to rock and perhaps to tip at a different point than intended. The consequences for Yorkshire's wagons was a supply of coal to a foundry's furnaces (or a loss of coal for an unintended tip); for some Earth system tipping points, the consequences might be a large economic shock reflected in the cost of adapting to or suffering from the new state of the tipping element.

*Lenton et al.* [2008]'s climatic definition retained the second element of Gladwell's definition (a large change resulting from a small change). It softened the first element – like contagions, climatic tipping points involve positive feedbacks, but not necessarily the specific, network-based feedbacks that characterize contagions. Importantly, the climatic definition drops

Gladwell's third element altogether: state shifts in climatic tipping elements need not occur rapidly, but can instead play out over an extended period of time. While the meaning of 'rapidity' depends on scale, the difference between 'committed' and 'realized' change gives rise to a scale-invariant distinction. This distinction has an analogy in chemical systems, where kinetic rate limitation may cause a system's transformation to lag significantly behind a shift in its equilibrium state. For some climatic tipping elements, a large *committed* change can occur rapidly as a result of a small change in forcing, though rate limitation may lead the *realized* change to occur slowly. For example, see Figure 2g-h, where the dashed lines indicate committed change, and the solid lines indicate realized change.

Global mean sea-level (GMSL) rise illustrates this distinction. *Kopp et al.* [2014] estimated that, under the moderate-emissions Representative Concentration Pathway (RCP) 4.5 (median projected warming of ~2.8°C above pre-Industrial by 2100), median projected *realized* GMSL rise for 2100 is about 0.6 m. By contrast, *Levermann et al.* [2013] estimated that the *committed* GMSL rise for 2.8°C of warming is about 6.4 m, and paleoclimatic evidence suggests that the committed rise may be higher still [*Dutton et al.*, 2015] – but this commitment may take centuries or millennia to be realized. Although there is considerable uncertainty associated with projections of both realized and committed change in response to a given temperature forcing, the distinction between these two changes is broadly recognized in the sea-level literature and is tied primarily to lags within the ice sheet and deep ocean responses.

The difference between committed and realized change is significant from a human perspective for three reasons. First, in some systems, the frictions that lead to a separation of realized and committed change may push the consequences of a tipping point beyond the time horizon of socio-economic relevance. For example, paleoclimatic evidence from the Last Interglacial suggests that the committed response to 2°C global warming above pre-Industrial temperatures may be about 6–9 m of GMSL rise [*Kopp et al.*, 2009], but if these 6–9 m were to take millennia to be realized, humans and ecosystems might readily adapt to them at minimal cost – in which case they will be of little socio-economic consequence. (In Figure 2g, the rate at which the system shifts states is considerably slower than in Figure 2e).

Second, some tipping elements exhibit early warning signs, such as a critical slowing down in rate of variability and an increase in the magnitude of variability, as they approach a critical threshold [*Dakos et al.*, 2008; *Scheffer et al.*, 2009, 2012; *Lenton*, 2011]. For a forcing that is rapid relative to a system's timescale of variability, whether these warning signs are detectable will relate to the lag between commitment and realization. If there is little or no lag, there will be no opportunity for warning signs. Conversely, if there is a long lag that reflects a generally slow system response time, any early warning signs of a committed state shift may be too slow to identify. Useful early warning signs require both that the speed with which a state shift is realized is slow relative to the system's timescale of variability, so that changes in rate or magnitude of variability have time to manifest before the realized state shift, and that the timescale of variability be sufficiently fast that such changes are detectable.

Third, provided the committed state shift can be detected, lags between realized and committed changes may allow for interventions, either by reversing the forcing that originally tipped the system or by introducing a different forcing. For example, for some temperature-triggered tipping points, bringing temperature back down below the tipping point quickly enough – before the previously committed change is fully realized – might avert some or all of that change. Reducing warming rapidly to 0°C after peaking at 2.8°C might avoid much of the committed sea-level rise, but because of hysteresis, there is no guarantee. For instance, in Figure 2g-h, the state shift continues well after the forcing starts to decrease. The desire to avoid committed state shifts has been an undercurrent in much of the discourse regarding 'albedo modification' or 'solar radiation management' (SRM), a climate engineering technique that can decrease global mean temperature [*Bickel and Agrawal*, 2012; *Moreno-Cruz and Keith*, 2012; *Keith*, 2013; *Irvine et al.*, 2014; *Heutel et al.*, 2015]. Some researchers suggest that SRM might in certain contexts reduce committed changes or slow their realization in a tipping element such as an ice sheet [*Irvine et al.*, 2009; *Applegate and Keller*, 2015]; others argue that, by the time a committed state shift is recognized, significant change may be long locked in [*Lenton*, 2011; *Barrett et al.*, 2014; *Markusson et al.*, 2014; *Sillmann et al.*, 2015].

Alternatively, though tipping points are generally defined with respect to a single type of forcing (e.g., temperature) while holding other forcings constant, a different type of forcing might change the commitment. For example, a cart sliding down a steep hill may be dynamically committed to landing at the bottom of the hill. If there is sufficient friction, however, there may be time to apply an external force to guide the cart onto a less steep slope and avoid the full, originally committed descent. Similarly, a permanent system for pumping water into the interior of the Antarctic ice sheet might – at enormous energetic cost – redefine the current sea-level rise commitment [*Frieler et al.*, 2016].

Climate economists often link the economic consequences of tipping thresholds to sometimes elusively defined 'economic catastrophes.' For example, in *Nordhaus and Boyer* [2000]'s DICE-99 integrated assessment model (IAM), an estimated willingness to pay to avoid such catastrophes – 1.0% of output-weighted global GDP at 2.5°C and 6.9% of output-weighted global GDP at 6°C – constituted about two-thirds of the total estimated economic damage caused by climate change. That rough proportion of catastrophic to total damages remained through the 2007 and 2010 revisions to DICE [*Nordhaus*, 2007].

*Nordhaus and Boyer* [2000]'s numbers were derived from an earlier survey of a group of economists and scientists [*Nordhaus*, 1994], which asked respondents to evaluate the probability that different climate pathways would cause a Great Depression-scale "high-consequence [economic] outcome," defined as a 25% loss of global income [*Nordhaus*, 1994]. For a 3°C rise in global mean temperature in 2090, the estimated probabilities ranged from 0 to 30%, with a mean probability of 0.6% and a median probability of 5%. Natural scientists averaged 12%, while non-environmental economists averaged 0.4%. Noting growing concern about the possibility of what they now called 'catastrophic' impacts, *Nordhaus and Boyer* [2000] doubled the mean estimates of *Nordhaus* [1994] without further explanation and remapped the 3°C probability to 2.5°C. They estimated that, due to risk aversion, the willingness to pay to avoid such 'catastrophic' losses would be ~2-3 times the expected loss.

*Nordhaus* [1994] neither asked his experts nor suggested to them how warming would cause a Great Depression-scale crisis. *Nordhaus and Boyer* [2000], however, offered a list of possible triggers for such 'economic catastrophes.' This list was rooted not in the economy, but in the physical climate, and bears a close resemblance to lists of potential climatic tipping points. "There are many concerns about catastrophic impacts of climate change," they wrote. "Among the potential severe events are a sharp rise in sea level, shifting monsoons, a runaway greenhouse effect, collapse of the West Antarctic Ice Sheet, and changing ocean currents that would have a major cooling effect on some subregions" [*Nordhaus and Boyer*, 2000, p. 87]. They did not further explore how such physical triggers might lead to a global Great Depression-scale economic loss.

## 2 Terminology

Muddled terminology has thus characterized the past two and a half decades of research on 'catastrophic', 'singular', 'abrupt', or 'tipping' impacts of climate change. We distinguish between *tipping elements*, *tipping points*, and *climate-economic shocks*. We apply the Lentonian definition of *tipping elements*: systems with critical thresholds, beyond which small perturbations in forcing can – as a result of positive feedbacks – lead to large, non-linear committed shifts in state. In our terminology, these tipping elements need not be limited to large-scale elements of the Earth system, but can also apply to social or engineered systems.

We suggest a use of the term *tipping point* that is more restrictive than the term 'critical threshold.' Because of the demonstrable influence of Gladwell's popularization, we suspect that non-specialists have something approximating the Gladwellian definition in mind when they hear the term, and that they therefore may be surprised to hear of 'tipping points' in which crossing a critical threshold may commit to a change that may not be realized rapidly but instead over centuries or millennia. We accordingly recommend that the term *tipping point* be reserved for Gladwellian critical thresholds, which we define as the critical thresholds exhibited by tipping elements with no significant lag between commitment and realization, and recommend that the generic term 'critical threshold' be used more broadly.

The term 'economic catastrophe' is charged and poorly defined. Even though *Nordhaus* [1994] used a benchmark of a 25% decline in global income, which he called "the economic equivalent of the Great Depression", it is unclear whether this statistic is relative to pre-decline output or to potential output. While GDP in the U.S. and Canada declined by 29% from 1929 to 1933, summed data from the 67 countries with available statistics show a 10% 'global' decline in GDP from 1929 to 1932 (followed by an increase in 1933) [*The Maddison Project*, 2013]. In the 61 countries with data available in 1921, total GDP was 27% lower in 1932 than it would have been had GDP growth continued at the 1921-1929 average annual rate. Two alternative interpretations of Nordhaus's benchmark would thus be 'a decline in global GDP comparable to that in North America during the Great Depression' or 'a decline in global GDP, relative to potential output, comparable to that during the Great Depression.' Different surveyed experts may have  interpreted this statement differently.

We recommend instead that the term *climate-economic shock* be used to refer to large economic losses triggered, directly or indirectly, by climate change. The qualifier 'large,' although not benchmarked, is intended to exclude shocks not readily identifiable above the natural background of weather shocks, while the use of the term 'shock' reflects the uncertainty around their occurrence. Climate-economic shocks may result from state shifts in climatic tipping elements (as suggested in the IAMs of *Nordhaus* [2000] and *Hope* [2013]) or climatically sensitive social tipping elements, but they may also result from the gradual effects of climate change. Moreover, not all state shifts in tipping elements – especially in non-Gladwellian tipping elements in which committed changes are realized over timescales long enough for ready human and ecological adaptation or other possible interventions – may lead to climate-economic shocks.

## 3 Climatic tipping elements

*Nordhaus and Boyer* [2000]'s list highlighted some potential state shifts in climatic tipping elements: a rapid sea-level rise driven by West Antarctic Ice Sheet collapse or by other sources; shifts in weather patterns like the Indian Summer Monsoon or the West African Monsoon; shutdown of the Atlantic Meridional Overturning Circulation (AMOC); or 'runaway' increases in climate sensitivity. Other studies have highlighted additional candidates, ranging from a collapse of Arctic summer sea ice, to ecological regime shifts in the Amazon or the Sahel, to a massive release of carbon from permafrost or seafloor methane hydrates [*Lenton et al.*, 2008; *National Research Council*, 2013]. Modeling studies have revealed the potential for an atmospheric superrotation threshold that rapidly increases climate sensitivity by changing planetary cloudiness [*Caballero and Huber*, 2013; *Pierrehumbert*, 2013], Arctic winter sea ice collapse, and an abrupt drop in the volume of snow on the Tibetan Plateau [*Drijfhout et al.*, 2015]. Some of these candidate tipping elements may exhibit Gladwellian tipping points, and so we categorize them as potentially Gladwellian tipping elements; others exhibit a non-Gladwellian disconnect between committed and realized change, and so we categorize them as non-Gladwellian (see Table 1 for an illustrative list). Potentially Gladwellian tipping elements generally involve components of the Earth systems with response timescales on the order of a decade or less. These components – many of which play an important role in 'fast' climate feedbacks [e.g., *PALAEOSENS Project*, 2012] – include the atmosphere, the surface ocean, and sea ice; slower responding components, such as ice sheets or the deep ocean, introduce significant lags between commitment to a state shift and its realization. We describe some examples in greater depth below.

### 3.1 Potentially Gladwellian tipping elements

**AMOC** is perhaps the most iconic climatic tipping element, and paleoclimatic evidence suggests that it can indeed exhibit Gladwellian behavior. Reconstructions based on Atlantic basin carbon isotopic records [*Sarnthein et al.*, 1994] suggest that AMOC exhibited three modes during the last 30,000 years: a "normal" mode, similar to today, with vigorous North Atlantic Deep Water (NADW) forming in the Nordic Sea; a "slowed down" mode, with reduced NADW formed to the south of Iceland; and a "collapsed" mode, in which NADW formation ceased and Antarctic Bottom Water filled the Atlantic basin [*Alley and Clark*, 1999; *Rahmstorf*, 2002]. Abrupt Dansgaard-Oeschger (D-O) climatic oscillations during the last glacial period [*Dansgaard et al.*, 1982] are thought to be expressions of transition between the AMOC modes [*Buizert and Schmittner*, 2015]. Geochemical proxies from the Bermuda Rise show that slowdowns and collapses of the AMOC occurred during Heinrich events, when icebergs discharge into the North Atlantic freshened the surface water, with no evidence for lags [*Böhm et al.*, 2015]. While D-O and Heinrich events occur during glacial periods, AMOC has also exhibited instability during past interglacials. Based on sedimentary carbon isotopes from Eirik Drift, near Greenland, *Galaasen et al.* [2014] found abrupt, multicentennial reduction in NADW during the Last Interglacial (~116–128 thousand years ago), with the largest and longest NADW reduction immediately following an outburst flood and a marked surface freshening event.

AMOC stability has also been investigated with numerical models of various complexities [*Stommel*, 1961; *Bryan*, 1986; *Hawkins et al.*, 2011; *Weaver et al.*, 2012]. In models, the existence of stable multiple equilibria for AMOC relies on the salt-advection feedback [*Stommel*, 1961]: a weakened AMOC decreases salinity in the North Atlantic and thus reduces the rate of deep water formation, which leads to further AMOC slowdown. In hosing experiments, which add freshwater to the North Atlantic, AMOC transitions to a collapsed mode once the input freshwater exceeds a threshold (~0.1-0.5 Sverdrups; [*Rahmstorf*, 2005]). AMOC also exhibits hysteresis: in order to restore the circulation once the system collapses, freshwater forcing must be reduced below a threshold that is smaller than the threshold originally required to cause the collapse [*Rahmstorf*, 1996, 2005; *Hawkins et al.*, 2011]. Although Southern Ocean warming associated with increased radiative forcing may increase the stability of AMOC and the threshold for freshwater-induced AMOC collapse [*Buizert and Schmittner*, 2015], North Atlantic surface warming reduces the density of water there, which inhibits deep water formation and weakens the AMOC in CMIP5 models [*Mikolajewicz and Voss*, 2000; *Gregory et al.*, 2005; *National Research Council*, 2013]. Therefore, AMOC strength is projected to decrease by the end of the century even in the low-emissions RCP 2.6 pathway; the magnitude of the decrease ranges between 5%–40% under RCP 4.5 and 15%–60% under RCP 8.5 [*Weaver et al.*, 2012; *Cheng et al.*, 2013]. A full collapse of AMOC in the 21st century is absent in most models except for FIO-ESM [*Drijfhout et al.*, 2015]. Based on model projections, *National Research Council* [2013] concluded that an abrupt slowdown or collapse of the AMOC due to anthropogenic forcing is very unlikely to occur in the 21st century.

AMOC transports a large amount of heat northward (up to $1 \times 10^{15}$ W) [*Ganachaud and Wunsch*, 2000], so its changes have large impacts on regional temperature and precipitation in the North Atlantic and the Northern Hemisphere. The slowdown of AMOC may be an important contributing factor for the "global warming hole" in the North Atlantic south of Greenland [*Drijfhout et al.*, 2012; *Woollings et al.*, 2012b; *Rahmstorf et al.*, 2015], where surface temperature displays a cooling trend that contrasts with the global warming trend [*Rahmstorf et al.*, 2015]. Hosing experiments in coupled climate models [*Laurian et al.*, 2009; *Drijfhout*, 2010, 2015; *Jackson et al.*, 2015] suggest that full collapse of the AMOC results in a significant northern hemisphere cooling (~2-5°C in the subtropical gyre [*Jackson et al.*, 2015]) and modest southern hemisphere warming (<1°C in most regions; [*Jackson et al.*, 2015]). These asymmetric temperature changes led to a decrease in global mean temperature (~0.7°C in the ECHAM5/MPI-OM model [*Laurian et al.*, 2009]) that offsets about ~15-20 years of global warming and causes a ~40-50 year warming hiatus [*Drijfhout*, 2015]. AMOC slowdown could also impact atmospheric circulations in several ways, including strengthening the North Atlantic storm track or reducing the frequency of 'polar low' cyclones in the subpolar North Atlantic [*Woollings et al.*, 2012a, 2012b]. It could also lead to poleward expansion of Hadley cells [*Drijfhout*, 2010], which may decrease precipitation over the Northern Hemispheric mid-latitudes and shift the Intertropical Convergence Zone southward [*Jackson et al.*, 2015]. The general cooling and atmospheric circulation changes would result in weaker peak river flows and vegetation productivity, which may raise issues of water availability and crop production [*Jackson et al.*, 2015]. Slowdown of the AMOC would also cause large and potentially rapid dynamic sea level changes; a collapse could raise sea level along the North American Atlantic coast by as much as ~0.5 m [*Gregory and Lowe*, 2000; *Levermann et al.*, 2005].

Both observational and modeling evidence suggests that, if it is a true tipping element, **Arctic sea ice** exhibits Gladwellian behavior. Observations show a significant decrease in sea-ice area in response to recent warming, with linear trends of $-54.6 \pm 3.7 \times 10^3$ km²/year annually and $-89.0 \pm 9.5 \times 10^3$ km²/year in September between 1978 and 2013 [*Stroeve et al.*, 2011, 2012; *Simmonds*, 2015]. Using the ECHAM5/MPI-OM GCM, *Li et al.* [2013] found no lag between changes in Northern Hemispheric temperature and changes Arctic sea ice area. Arctic sea ice also has the potential to be a tipping element, as sea-ice loss can be amplified by feedbacks involving ice albedo, the warming effects of convective clouds, the open-water formation efficiency of thin ice, and the increased temperature responsiveness of thinner, younger ice [*Drijfhout et al.*, 2015].

Energy balance models and single column models [*Rose and Marshall*, 2009; *Björk et al.*, 2013] suggest that Arctic sea ice has a critical threshold below which rapid ice-cover shrinkage will occur and lead the Arctic to be seasonally ice free [*Notz*, 2009]. *Bjork et al.* [2013] placed this threshold at an annual-mean ice thickness of 1.7–2.0 m, but the tipping behavior of Arctic sea ice simulated in comprehensive GCMs is still controversial [*Armour et al.*, 2011; *Ridley et al.*, 2012; *Li et al.*, 2013; *Wagner and Eisenman*, 2015]. In ECHAM5/MPI-OM, *Li et al.* [2013] found that summer sea-ice area declines linearly in response to increased $CO_2$ concentration, while winter sea-ice area shows a rapid transition (at a temperature higher than that which causes a complete loss of summer sea ice) to a nearly ice-free state. On the other hand, an abrupt transition of Arctic sea ice is not found in any season in CCSM3 [*Armour et al.*, 2011] or HadCM3 [*Ridley et al.*, 2012]. There is also no consensus on the persistence or hysteresis of Arctic sea decline in comprehensive GCMs [*Armour et al.*, 2011; *Li et al.*, 2013; *Wagner and Eisenman*, 2015]. In CMIP5 climate models, increasing greenhouse gas are projected to drive a steady decrease in Arctic sea ice, with a possibility of ice-free summer within a few decades [*Stroeve et al.*, 2012; *Overland and Wang*, 2013; *Overland et al.*, 2014]. Five CMIP5 models show an abrupt winter Arctic sea-ice abruptly collapse in the 22[nd] century in RCP8.5 simulations [*Drijfhout et al.*, 2015]. Overall, the evidence that winter Arctic sea ice is a tipping element is stronger than for summer Arctic sea ice. However, *Bathiany et al.* [2016] argue that the abruptness of winter Arctic sea ice collapse can be explained by a threshold without positive feedbacks, simply from a basinwide failure to cool sufficiently to allow ice formation; if they are correct, neither winter nor summer Arctic sea ice may be tipping elements.

A rapid decrease in Arctic sea ice could have far-reaching consequences. Arctic sea ice and the ice-albedo feedback are an important contributor to Arctic amplification, the phenomenon that surface warming over the Arctic is more rapid than at lower latitudes. Arctic amplification may slow down the mid-latitude jet streams, shift storm tracks over the North Atlantic, and increase the vertical propagation of energy into the stratosphere, which may lead to more frequent extreme weather events across the Northern Hemisphere mid-latitudes [*Francis and Vavrus*, 2012, 2015; *Cohen et al.*, 2014; *Tang et al.*, 2014]. However, some studies argue that the chance of mid-latitude cold extremes should decrease in response to future sea ice loss [*Hassanzadeh et al.*, 2014; *Screen et al.*, 2014]. In addition, this rapid warming over the Arctic due to sea-ice reduction may consequently increase the emission of methane from high-latitude wetland soils; *Parmentier et al.* [2015] estimated that methane emission over high-latitudes for 2005–2010 are averagely ~1.7 Tg yr[−1] higher than that during 1981–1990 due to a sea ice-induced warming in fall.

The **West African monsoon (WAM)** may be another example of a Gladwellian tipping element. WAM contributes the bulk of Sahelian summer rainfall [*Dong and Sutton*, 2015]. Paleoclimatic evidence suggests that the bistability of the WAM is characterized by alternating status between long-lasting (decades to centuries) episodes of dry and wet conditions [*Shanahan et al.*, 2009]. Both paleoclimatic reconstructions [*Asmerom et al.*, 2013] and GCM simulations [*Giannini et al.*, 2003; *Hoerling et al.*, 2006; *Martin et al.*, 2014] suggests that the switch between these two quasi-stable states is driven by changing sea surface temperatures (SST) around Africa. Sediment-core and tree-ring reconstructions indicate that the WAM variability is coherent and in phase with the Atlantic SST variability at multidecadal time scale [*Shanahan et al.*, 2009], suggesting Gladwellian behavior. The transition of WAM to its strong phase is associated with a wind-evaporation-SST positive feedback [*Xie*, 1999]: warming in SST in the North Atlantic relative to the South Atlantic drives stronger westerly winds and enhances the WAM, which reduces surface evaporation north of the equator and enhances it in the south, amplifying the interhemispheric SST gradient. In addition, a switch to strong WAM may also rely on the Saharan water vapor–temperature feedback [*Evan et al.*, 2015]: long-wave radiation of surface water vapor raises the surface air temperature, which increases the low-level moisture convergence around the Saharan heat low.

An enhanced WAM driven by anthropogenic greenhouse gas and aerosol forcing may have led to the substantial recovery of Sahel rainfall since the 1980s [*Dong and Sutton*, 2015]. Under RCP 8.5, about 80% of CMIP5 models agree on a modest drying around 20% over the westernmost Sahel (15°–5°W), while about 75% of models agree on an increase in precipitation over the Sahel between 0°and 30°E, with a large spread on the amplitude [*Roehrig et al.*, 2013]. However, the CMIP5 models may underestimate the monsoon decadal variability, due to strong biases in simulated SST [*Roehrig et al.*, 2013]. One model, BNU-ESM, found an abrupt increase in Sahel vegetation cover around 2050 in the RCP8.5 simulation [*Drijfhout et al.*, 2015]. The projected WAM strengthening is related to a robust amplification of warming over the Sahel by about 10%–50% over the global mean [*Roehrig et al.*, 2013]. Furthermore, the warming pattern induced by enhanced WAM facilitates the development of African easterly waves (AEWs), which are westward-propagating weather disturbances over North Africa during summer. In RCP8.5 simulations between 2075 and 2100, the occurrence frequency increased with a multi-model average of 39% for intense AEWs and of 72% for extremely intense AEWs along the Sahel−Sahara border [*Skinner and Diffenbaugh*, 2014]. The elevated AEW activity could further increase the Sahel rainfall, and strengthen Sahara dust transportation over Africa and Atlantic [*Skinner and Diffenbaugh*, 2014].

### 3.2 Non-Gladwellian tipping elements

**Ice sheet** melt provides clear examples of non-Gladwellian behavior. For the Greenland Ice Sheet, for example, feedbacks between ice sheet topography and atmospheric dynamics and between ice area and albedo give rise to multiple stable states [*Ridley et al.*, 2009; *Robinson et al.*, 2012; *Levermann et al.*, 2013]. *Robinson et al.* [2012]'s coupled ice-sheet/regional climate model indicated that, at a temperature of 1°C above pre-Industrial, the stable states are at 100%, 60%, and 20% of present ice volume. At 1.6°C, however, their model produced only one stable configuration, at ~15% of the Greenland ice sheet's present volume; thus, 1.6°C warming would represent a commitment to ~6 m of sea-level rise from the Greenland Ice Sheet. The rate of ice sheet mass loss is, however, limited by the flux at the ice sheet margins [e.g., *Pfeffer et al.*, 2008], leading to a disconnect between committed and realized change that could persist for millennia, particularly for levels of warming near the threshold [*Applegate et al.*, 2015].

In Antarctica, where about 23 m sea-level equivalent of ice sits vulnerably with its base below sea level [*Fretwell et al.*, 2013], ice sheet mass loss is dominated by ocean/ice sheet/ice shelf interactions. For parts of the ice sheet sitting on reverse bed slopes, which shallow outward, a positive feedback sets up the marine ice sheet instability [*Schoof*, 2007; *Gomez et al.*, 2010; *Ritz et al.*, 2015]: as the ice retreats, it thickens vertically, increasing ice sheet discharge and thus the rate of retreat. Such an instability appears to be occurring in multiple outlet glaciers of the Amundsen Sea Embayment sector of the West Antarctic Ice Sheet, creating a sea-level rise commitment that could equal much or all of the 1.2 m sea-level equivalent in this sector [*Joughin et al.*, 2014; *Rignot et al.*, 2014]. In Wilkes Basin in East Antarctica, the instability threshold has not yet been crossed, but removal of an 'ice plug' containing ~8 cm sea-level equivalent could create a 3-4 m sea-level rise commitment [*Mengel and Levermann*, 2014]. Ice shelf buttressing can inhibit marine ice sheet instability, however, so some marine-based sectors may not exhibit threshold behavior but instead respond nearly linearly to sub-shelf temperature [*Mengel et al.*, 2016]. On the other hand, most ice sheet models do not include ice cliff collapse and hydrofracturing, which destabilize ice shelves and may greatly increase the rate of ice sheet mass loss [*Pollard et al.*, 2015; *DeConto and Pollard*, 2016]. Overall, the Antarctic ice sheet exhibits a clear separation between realized and committed change, and individual sectors of the ice sheet can exhibit threshold behavior associated with marine ice sheet instability. It is unclear, however, whether threshold behavior occurs at the aggregate level of the ice sheet as a whole [*Levermann et al.*, 2013].

Model evidence also suggests that **large-scale ecosystems** can exhibit a non-Gladwellian lag between committed and realized state shifts. Using the HadCM3LC climate/carbon cycle model, which was one of the first to include a dynamic vegetation component, *Jones et al.* [2009] found a precipitous committed collapse in Amazon forest cover between 1°C and 3°C warming, even though the realized loss when their simulation reached 3°C was minimal. Similarly, they found a committed tripling of boreal forest cover at ~4°C warming, even though the realized expansion was minimal. These results were not consistent across models [*Cox et al.*, 2013]; among model participating in the Coupled Climate Carbon Cycle Model Intercomparison Project (C4MIP) exercise, only the HadCM3 models showed a net reduction in tropical land carbon over the 21$^{st}$ century. Conditioning the C4MIP models on the observed inter-annual variations in $CO_2$ growth rate over 1960-2010 suggested that the tropical forests in the HadCM3 models are overly sensitive to warming [*Cox et al.*, 2013], but to our knowledge no other models have been used to investigate the distinction between committed and realized ecosystem change.

### 3.3 Common worrisome traits of candidate climatic tipping elements

Some candidate tipping elements exhibit full-fledged Gladwellian tipping points, others exhibit non-Gladwellian critical thresholds, and some, upon further investigation, may turn out not to be tipping elements at all. But most of these ambiguous candidate tipping elements unambiguously exhibit other worrisome traits. First, almost all are characterized by deep uncertainty [*Kasperson*, 2008; *Heal and Millner*, 2014; *Convery and Wagner*, 2015], meaning that there are multiple plausible probability distributions that could be constructed for how likely they are to occur, how fast they will occur, and what their consequences could be. This deep uncertainty complicates efforts to devise policies to minimize negative consequences. Second, and closely related, they are absent from many of the coupled climate models used to project future changes: some models will generate the tipping behavior, some will not, and some are missing the element of the Earth system that could generate them. Third, many are rate-limited and so exhibit a committed change significantly larger than the initially realized change. Fourth, many exhibit hysteresis, so reversing a change in the system may require a larger forcing and/or more time than causing the change in the first place.

For example, the reservoir of **soil organic carbon in the Arctic permafrost** may or may not abruptly 'tip'. Arctic permafrost holds at least ~1300-1600 Gt C [*Schuur et al.*, 2015]; as the Arctic warms, microbes will transform this organic carbon into $CO_2$ and $CH_4$. If all of this carbon were instantaneously released as $CO_2$, it would likely cause a global mean warming of ~1-3°C. Thus, there could be a positive feedback large enough to create a tipping point: the release of carbon could warm the planet enough to significantly accelerate the rate of carbon release. Moreover, laboratory incubations show that some organic-rich permafrost soils can decompose rapidly, with up to ~7% of organic C being lost in the first unfrozen year, [*Schädel et al.*, 2014] and that abrupt permafrost thaw is common phenomenon in parts of the Arctic [e.g., *Jorgenson et al.*, 2006].

Yet the real-world rate of permafrost carbon release is limited by the annual freeze-thaw cycle, by the rate of thermal diffusion into the deep permafrost, by the creation of new biomass, and by oxygen availability in water-logged soils [*Schuur et al.*, 2015]. (Anaerobic decomposition creates methane, but current estimates indicate the higher warming potential of methane is insufficient to offset the decreased decomposition rate in oxygen-depleted soils.) A 17-author expert assessment estimated that ~5-15% of permafrost C is vulnerable to decomposition in the 21st century, and model projections suggested a somewhat larger share vulnerable to decomposition in the 22nd and 23rd centuries [*Schuur et al.*, 2015]. If this assessment is correct, then the permafrost feedback on warming will be too small to be the principle driver of permafrost melt; there will be no threshold beyond which the momentum of escalating permafrost emissions carries the world to a permafrost-free state.

Nonetheless, Arctic permafrost carbon still shares the four other traits common to many tipping elements – deep uncertainty, incomplete treatment in climate models, a separation between committed and realized change, and hysteresis. It thus revises our understanding of the coupled climate/carbon cycle system.

A significant body of work over the last decade has shown that global mean warming increases approximately linearly with cumulative carbon dioxide emissions, and that global mean temperature is stable for centuries after emissions stop [e.g., *Matthews and Caldeira*, 2008; *Allen et al.*, 2009; *Solomon et al.*, 2009]. The IPCC concluded that "the principal driver of long-term warming is total emissions of $CO_2$ and the two quantities are approximately linearly related" [*Collins et al.*, 2013, p.1033]. These conclusions gave rise to the 'carbon budget' paradigm now common in policy discourse. But the models that led to these conclusions did not include permafrost carbon. Some newer Earth system models have incorporated the permafrost carbon feedback, at least in part. *Schaefer et al.* [2014]'s synthesis found a mean initial carbon pool among thirteen published studies of ~800 Gt C, roughly half the observational estimate. These models found that the permafrost carbon feedback lags anthropogenic emissions, giving rise to a difference between committed and realized emissions similar to that for ice sheet melt. As a consequence, the approximation that warming is proportional to cumulative anthropogenic $CO_2$ emissions fails, and warming may continue after human emissions stop [*MacDougall et al.*, 2012].

Using the UVic ESCM, an Earth system model of intermediate complexity (EMIC), *MacDougall et al.* [2015] found that incorporating the permafrost carbon feedback led to a ~10% reduction in the carbon budget for 2°C warming. The permafrost carbon feedback similarly led to an increase of ~10% in the amount of net anthropogenic carbon removal needed to restore a 2°C warming after overshooting to 3.2°C. Using another EMIC, IAP RAS CM, *Eliseev et al.*

[2013] demonstrated hysteresis in the permafrost system, showing that permafrost melted faster in a warming world than it regrew in a cooling world at the same temperature.

3.4 Integrated assessment of climatic tipping elements

Regardless of whether all proposed tipping elements do in fact tip, their potential state shifts are hazards; it is therefore worthwhile to identify and assess climate-economic shocks that they might cause. Indeed, since abstract 'catastrophic' impacts dominate damage estimates in two of the most commonly used benefit-cost IAMs, DICE and PAGE [*Hope*, 2013], improving their representation in IAMs may be critical to more accurate estimates of the cost of climate change [*Revesz et al.*, 2014].

*Lenton and Ciscar* [2013] outlined one strategy for a stylized improvement of climatic tipping element representation in benefit-cost IAMs. A number of subsequent studies have focused on the welfare costs of uncalibrated or idealized 'tipping points'. *Lemoine and Traeger* [2014] examine uncalibrated, instantaneous changes in climate sensitivity or carbon sinks in a variant of the DICE model; *Lemoine and Traeger* [2016] expanded this study to include 'tipping points' that affect the damage function directly and also show interactions between 'tipping points.' *Daniel et al.* [2015] decompose the impact of risk aversion to 'tipping points' from climate damages. *Lontzek et al.* [2015] consider a single idealized critical threshold in the damage function with a range of transition scales and final damage levels. *Cai et al.* [2016] implemented five interacting tipping elements (AMOC, West Antarctic Ice Sheet, Greenland ice sheet, the Amazon dieback, and El Niño-Southern Oscillation) with critical threshold probabilities calibrated to the expert elicitation study of *Kriegler et al.* [2009], transition timescales based on literature reviews, and economic consequences based upon the authors' intuition.

We propose an alternative strategy for improving representation of state shifts in tipping elements that draws upon more disaggregated models of climate change impacts [e.g., *Warszawski et al.*, 2014; *Houser et al.*, 2015]. For risk and impact assessment, potential climatic tipping elements can be categorized by the physical parameters they affect that influence human systems. (For simplicity, we will put aside the qualification 'potential' and refer to all potential tipping elements simply as 'tipping elements' for the remainder of this section.) Some tipping elements affect global warming, either through planetary albedo (e.g., Arctic sea ice) or the greenhouse effect (e.g., permafrost). Some tipping elements influence on regional temperature or precipitation (e.g., changes in AMOC or Arctic sea ice). Tipping elements involving polar ice sheets affect global mean sea level, and some tipping elements (e.g., AMOC) influence regional dynamic sea level. Finally, some tipping elements involve major ecosystems such as the Amazon, which may influence not only the greenhouse effect but also the regional availability of ecosystem services. Each category is amenable to a different risk assessment strategy.

Both benefit-cost IAMs like DICE and some newer frameworks based on empirical and process model of climate change impacts [e.g., *Houser et al.*, 2015] use simple climate models (SCMs) to project global mean temperature change. Tipping elements that affect global mean temperature could be incorporated into these SCMs. Although subject to well-known concerns about the damage functions in benefit-cost IAMs [e.g., *Revesz et al.*, 2014], two studies have already attempted this for permafrost carbon. *Hope and Schaefer* [2016] augmented the PAGE09 IAM [*Hope*, 2013] with time series of permafrost $CO_2$ and $CH_4$ emissions [*Schaefer et al.*, 2011]. They found that permafrost emissions increased the net present value cost of climate change by ~13%, consistent with the associated increased in cumulative $CO_2$ emissions. *González-Eguino and Neumann* [2016] conducted a similar study with DICE-2013R. Similar approaches could be used to study the global mean temperature effect of Arctic albedo changes due to sea-ice collapse or boreal forest expansion, of a superrotation-induced decrease in cloudiness and increase in climate sensitivity, or of rapid changes in the land or ocean carbon sinks.

Many tipping elements, however, have regional temperature and precipitation effects that extend beyond a scaling of regional changes with global mean temperature. Arctic sea ice loss decreases the pole-to-equator temperature gradient, which may affect the frequency of mid-latitude weather patterns [*Overland et al.*, 2015]. A reduction or collapse of AMOC has strong cooling effects in Europe, weaker cooling effects more broadly in the Northern Hemisphere, warming effects in the Southern Hemisphere, a southward shift in the Intertropical Convergence Zone, and sea-level rise in the North Atlantic [*Vellinga and Wood*, 2002; *Levermann et al.*, 2013; *Jackson et al.*, 2015]. A change in ENSO frequency or intensity would have temperature and precipitation effects around the world [*Power et al.*, 2013; *Cai et al.*, 2014; *Latif et al.*, 2015; *Yuan et al.*, 2015]. GCMs must be used to identify the spatial and temporal patterns associated with state shifts in such tipping elements. Either direct GCM output or the extracted patterns can then be combined with empirical [e.g., *Dell et al.*, 2014] or process [e.g., *Warszawski et al.*, 2014] models to estimate regional impacts and damage functions that go beyond temperature impacts.

A number of studies have attempted to assess the economic impact of AMOC collapse. Some [e.g., *Keller et al.*, 2000, 2004; *Mastrandrea and Schneider*, 2001] have linked simple models of AMOC stability to arbitrary perturbations of the damage function of a simple benefit-cost IAM like DICE. *Link and Tol* [2010] combined the spatial pattern of AMOC collapse from an experiment with the HadCM3 GCM with the impact functions from the FUND 2.8 IAM, which has temperature-driven damage functions for each of 16 regions and six impact categories. They found a small global effect (a ~0.1% reduction in global GDP in 2100), though larger negative effects (up to ~4% of GDP) in a few, mostly high-latitude countries. *Kuhlbrodt et al.* [2009] linked the temperature, precipitation, cloudiness and pressure precipitation projections downscaled from the Climber-3a intermediate-complexity Earth system model to the LPJmL dynamic vegetation model to assess the effect of an AMOC shutdown on European crop production; they found a fairly limited effect. *Link and Tol* [2009] linked the ocean temperature and AMOC projections of Climber-3a to a bioeconomic model of the Barents Sea cod fishery. They found that the direct effect of AMOC weakening on survival rates could lead to a fisheries collapse. Similar studies could be conducted for other tipping elements with regional climatic effects, and the empirical climate impact functions increasingly emerging in the econometric literature could be leveraged to link regional climate changes to their socio-economic consequences [e.g., *Dell et al.*, 2014].

State shifts in ice-sheet tipping elements are conceptually the easiest to incorporate into risk assessments. Their primary impact is to change rates of sea-level rise, and so they can be assessed in the same frameworks used to assess sea-level rise impacts more broadly. For example, *Nicholls et al.* [2008] forced the FUND model with West Antarctic Ice Sheet collapse, represented by 5 m of global-mean sea-level rise in a period as short as 100 years. For a century-timescale collapse, they found a ~40% drop in the length of the coastlines that it was benefit-cost optimal to protect, and a 15-fold increase in annual protection costs. *Diaz* [2015] and *Diaz and*

*Keller* [2016] extended DICE with a stochastic representation of West Antarctic Ice Sheet collapse, with collapse probabilities loosely calibrated against the expert elicitation study of *Bamber and Aspinall* [2013].

Arguably, the possibility of large-scale ice-sheet collapse is already built into some probabilistic sea-level rise projections [e.g., *Kopp et al.*, 2014], though these may understate the probability of this outcome [*DeConto and Pollard*, 2016]. For example, the *Kopp et al.* [2014] 99.9th percentile projections align with other estimates of the maximum physically plausible level of 21st century sea-level rise (~2.5 m) and require > 95 cm of rise driven by the Antarctic ice sheet in the 21st century. However, risk assessments based on these projections [*Houser et al.*, 2015] have not examined outcomes that far into the tail of the analysis. Deliberate inspection of such tail risks would provide a natural way of incorporating ice-sheet tipping elements into coastal risk assessments.

Ecological tipping elements may be the hardest to incorporate into economic risk assessments, particularly at a global scale. The difficulty reflects the state of the field of ecosystem services valuation. Generally, assessments of ecosystem services are either narrowly focused [e.g., *Jenkins et al.*, 2010] or fairly vague [e.g., *Costanza et al.*, 1997; *de Groot et al.*, 2012]. The easiest risks of large-scale ecosystem changes to assess may be those that feedback onto global radiative balance, through either changes in the carbon cycle or changes in land surface albedo.

In summary, potential climate-economic shocks associated with state shifts in different tipping elements can be independently assessed, for example, by (1) linking simple climate model scenarios for elements that affect greenhouse gas concentrations or albedo to impact models and (2) incorporating regional spatial-temporal patterns associated with tipping elements with regional climatic effects to impact models. Some tipping elements are driven by natural-system variables only partially reflected in current Earth system models and thus require alternative approaches, such as (3) extending coastal risk assessments into the tail of sea-level rise probability distributions, which should incorporate the possibility of rapid ice-sheet melt; and (4) tallying the net costs of climate impacts on monetizable ecosystem services. However, tipping elements are not necessarily independent of one another [e.g., *Kriegler et al.*, 2009], so when estimating the economic risks of different elements, it is important to consider the correlations between different thresholds. For example, rapid Greenland ice mass loss increases the probability of AMOC collapse, which in turn leads to Southern Ocean warming and increases the probability of rapid West Antarctic Ice Sheet mass loss. In addition, the framework laid out above focuses on how a state shift in a tipping element may cause different individual types of impacts, but it is also important to consider how these impacts interact across sectors [e.g., *Warren*, 2011]. Severe negative (or positive) impacts on agriculture, for example, will affect food prices and lead to changes in economic structure. Similarly, migration away from coastlines could have either negative or positive economic effects. It is also important to keep in mind that one effect of a state shift may be to change the temporal and spatial correlations between extreme events, making once-rare 'black swan' [*Taleb*, 2007] alignments of extremes much more likely.

## 4 Climatically sensitive social tipping elements

Given the sociological origins of the modern 'tipping point' concept, it is perhaps surprising that climate change tipping point research has focused almost exclusively on tipping elements in natural systems, leaving climatically relevant social tipping elements mostly unexplored. Broader social change theory has, however, recognized that change is often not gradual. *Moser and Dilling* [2007] identified four typical stages of social transformation that together form a S-curve. In the "predevelopment" phase, the system is in one stable state. During "take off," it begins to accelerate towards a new state. The third stage, which they titled "breakthrough," is the tipping point, where the rate of change accelerates until it reaches the fourth stage, "stabilization" in a new state. Influencing this accelerating rate of social change is typically a positive feedback mechanism involving network diffusion, by which one actor (e.g., a household, a legislature, or a corporation) changes, and others – via social cues – exponentially follow suit. Social tipping points can influence climate change, and climate change can influence social tipping points. Similarly, economic shocks may both cause and be caused by social tipping points.

Like climatic tipping elements, social tipping elements both involve positive feedbacks and exhibit non-linear rates of change. This is a more restrictive definition of social tipping points than what is sometimes applied in the social scientific literature. While there are robust research fields that recognize thresholds in human-climate systems, they generally lack an explicit focus on positive feedbacks within social systems, instead considering the more generic case of systems with thresholds, like those shown in Figure 2b-d. For example, *Bardsley and Hugo* [2010, p.243] define a migration threshold as "a point at which the impacts of climate change are so severe or so frequent that the resilience of socio-ecological systems is breached, or that existing in situ adaptation options either fail or are perceived as inadequate, so that people make use of migration as an adaptation option in a manner that will fundamentally alter the form migration is taking". While migration can be a tipping element, this description does not capture the role of the feedbacks that make it such. Other researchers have found that as climate change impacts increase in magnitude, there is a threshold where individuals' responses change from being adaptive to maladaptive [*Niemeyer et al.*, 2005]. Researchers conducting interviews that presented worsening climate change scenarios saw an increase in the "adaptive response" of concern and action for a warming scenario (2.5°C). When this scenario shifted to a greater magnitude of climate change (5°C), the researchers observed an increase in apprehension and a decline in trust and perceptions that institutions and other social actors will respond. The researchers saw this as an increase in maladaptation and suggest that there is a threshold between 2.5°C and 5°C at which people become less adaptive. This study suggests a threshold in social behavior, but not a positive feedback mechanism through which it occurs. In addition, socio-ecological studies examine how human actions, climate change, and other factors trigger ecological regime shifts, but the tipping elements in these studies are ecological, not social [*Kinzig et al.*, 2006].

The term 'adaptation tipping point' [*Kwadijk et al.*, 2010; *Werners et al.*, 2013; *Koukoui et al.*, 2015] has been occasionally used in the literature to refer to sigmoidal changes in the frequency of extreme events above a threshold at which "the current management strategy will no longer be able to meet [its] objectives" [*Kwadijk et al.*, 2010, p. 730]. We advise against this use, as it is inconsistent with other uses of the term 'tipping point.' Although tipping points proper do produce sigmoidal changes, the sigmoidal changes associated with 'adaptation tipping points' do not arise from positive feedbacks and thus lack a crucial characteristic of true tipping points. They are simply a product of extreme value statistics, by which the number of exceedances of a threshold can grow approximately exponentially in response to a linear increase in the mean of a distribution. For example, the historical 1% average annual probability flood at the Battery tide gauge in New York City is ~1.8 m above mean high water. With 50 cm of sea-level rise, the expected number of such floods increases about 5 times; with 100 cm, by about 40 times, and with 150 cm, by about 2000 times [*Buchanan et al.*, 2016]. If current management strategies cannot cope with a '1-in-100 year' flood occurring with an annual probability >10%, the 'adaptation tipping point' would occur when sea-level rise exceeded ~70 cm.

Social tipping points can be beneficial, costly, or neutral to human welfare. Beneficial social tipping points increase societal resilience and reduce climate change damages via mitigation or adaptation. Harmful social tipping points are more likely to occur where there are low levels of societal resilience, under which societal risks increase because of failure to effectively adapt or mitigate. Potential social tipping elements that are relevant to integrated assessment of the costs of climate change include (1) public opinion and policy change, (2) technology and behavior adoption for adaptation or mitigation, (3) migration, and (4) civil conflict.

In **environmental policy**, the theory of 'punctuated equilibrium' explains intervals of long-term policy stasis and incrementalism, interrupted by abrupt changes to a new policy state [*Baumgartner et al.*, 2014]. These "explosive change[s] for a short while" lead to "the establishment of a new policy equilibrium" [*Baumgartner et al.*, 2014, p.61]. Unity in public opinion on an environmental issue, decreasing unity in opposition to policy change by negatively affected interest groups, and governmental openness to policy change can all influence policy change [*Shwom*, 2011]. These underlying factors are social phenomena that can experience positive feedbacks related to network influencer effects [*Watts and Dodds*, 2007].

Social scientists use "'sand-pile' models, 'tipping point' models, 'small world' models and other graph theoretic models, 'complex adaptive systems models', and models that produce punctuated dynamics via a hierarchy of time scales" [*Brock*, 2006, p.49] to capture the non-linearity of social change dynamics. Modeling environmental policy change suffers from the typical modeling challenges for complex social systems, including the identification problem (identifying what variables should be included in the models). Our review did not find any models of public opinion or politics of climate change, though there are a number of models of punctuated equilibrium and environmental policy change (see [*Repetto*, 2008] and [*Garmestani*, 2014] for reviews).

Climate change mitigation and adaptation require **adoption of new technologies and behaviors**. For example, to reduce greenhouse gas emissions, society will need to adopt energy-efficient and low-carbon energy technologies. Researchers have historically used logistic substitution models to represent energy technology uptake over time [e.g., *Marchetti and Nakicenovic*, 1979]. Energy technologies exhibit a gradual diffusion in the introductory phase, then experience exponential growth as the learning curve and up-scaling of a technology takes place (the tipping point), after which growth tapers off as the technology approaches saturation [*Grübler et al.*, 1999]. *Wilson* [2012] mapped a variety of energy production technologies and their capacity in the Netherlands over time and found that fossil fuel, nuclear, and renewable energy technologies all exhibit this general pattern, though length of time in each phase varies.

The "epidemic" model of technology diffusion [*Geroski*, 2000] posits that technology adoption is dependent upon information availability and experience spread through networks [*Rogers*, 2003]. For example, farmers in the Philippines used relatives and neighbors as a reliable source of information and credit to help adapt to drought [*Acosta-Michlik and Espaldon*, 2008]. Because proximity influences social networks and cues (e.g., a solar panel is visible on a neighbor's roof), spatial agent-based models have been used to model technology diffusion and adoption processes [*Berger*, 2001; *Noonan et al.*, 2015]. Many technological adaptations, such as raising houses, adopting air-conditioners, or modifying farming techniques, are innovations that experience these network feedbacks and will diffuse following a logistic curve.

Recent models of technology adoption have evolved beyond the diffusion curve to provide system-dynamic accounts. Approaches like energy innovation systems [*Gallagher et al.*, 2012] and socio-technical systems transitions [*Geels*, 2005; *Geels and Schot*, 2007] embed technology adoption within supply and demand dynamics, sources of technology change, the technology development cycle, innovation processes, and feedbacks between networks of actors and institutions. Learning and experience curves have been used as the basis for endogenous technological change in a number of formal models, but not without critique [see *Wilson*, 2012]. Clarke et al. [2008, p.413] highlight that "although learning-by-doing may be one of the factors underlying these curves, experience curves are a reflection of all factors that play into the change in technological performance and cost, including research and development, spillovers and economies of scale." In an analysis focusing on innovation processes of five renewable energy case studies, *Hekkert and Negro* [2009] found that actions that turn knowledge into business opportunities (entrepreneurial activities) rise when there is guidance that positively affect visibility and clarity of specific wants among technology users (e.g., a policy goal aim for a certain percentage of renewable energy in a future year). However, they found that technologies did not "tip" until there was market formation, such as the creation of a small protected market (via a subsidy or tax advantage) or a niche market for a specific application.

Another potential social tipping element involves **migration**, which is generally costly to the source region but can potentially be beneficial for recipient regions and for overall human welfare. Migration is an adaptation to new environmental conditions. Climate-related migration can involve forced displacement, where climate change has made it difficult to stay, or may take place in anticipation of future risks. Moving away from sub-Saharan Africa as desertification and water scarcity make survival more difficult is an example of climate displacement [*Bogardi and Warner*, 2009]. Migration can also be a response to social and economic destabilization that results from climate change impacts [*Warner et al.*, 2010]. These are examples of migration as a mass general social response, but, as we discuss below, there can also be positive feedbacks in migration systems.

Migration-systems approaches identify "push" factors that encourage or enable people to leave their home location and "pull" factors that encourage them towards a new location [*Mabogunje*, 1970; *Fawcett*, 1989; *Jennissen*, 2007]. Social networks are an important pull factor; once a migration stream has been initiated, it tends to grow, as the networks provide a positive feedback [*Boyd*, 1989; *Fawcett*, 1989; *Massey and Zenteno*, 1999]. *Massey and Zenteno* [1999] built a dynamic model of mass migration that account for feedbacks, providing the basis of an approach that could be used to model climate-related migration tipping points. As people migrate away in large numbers due to environmental change, a feedback can occur that changes a community's ability to adapt [*McLeman and Smit*, 2006].

**Civil conflict**, and in particular the 'conflict/development trap', can give rise to a more unambiguously costly tipping point. The conflict/development trap is a well-established positive feedback cycle, in which failure to develop increases the likelihood of civil conflict, which in turns decreases the ability to develop and increases the future risks of conflict [*Collier*, 2003]. This can give rise to a counter-development tipping point, as has arguably been seen in a number of strife-torn, low-income countries. Econometric results suggest that warmer temperatures can slow economic growth [*Burke et al.*, 2015] and increase the likelihood of civil conflict [*Hsiang and Burke*, 2013; *Hsiang et al.*, 2013]. Thus, climate change has the potential to make a conflict/development tipping point more likely, exacerbating the risk that nations may get trapped in a cycle of poverty and conflict that is difficult to break.

*Bentley et al.* [2014] investigated whether early warning signs, such as a critical slowing down or an increase in variability, are associated with social tipping points. They found that, historically, many social systems have undergone tipping points without exhibiting such warning signs. For example, they point to the case of small English banks, which increased in number at about 2.7%/year for 150 years and then suddenly declined in 1810 with no data indicating such a change was impending. While social scientists can identify potential social tipping elements and associated mechanisms, it is far more difficult to predict when they will occur, and studies must consider that a range of biophysical, social, cultural, political and economic factors influence social responses to climate changes [*Nuttall*, 2012].

Climatic tipping points may be one trigger for social tipping points. Another potential trigger are increases in the frequency of extreme events. Although the 'adaptation tipping points' of *Kwadijk* [2010] are not true tipping points, these dramatic increases in extreme event frequency may force true adaptation tipping points – state shifts in genuine social tipping elements, such as adaptive policies or behaviors, that lead to greater adaptation [*Pelling and Dill*, 2010]. For example, there may be a critical threshold in the frequency of days over 30°C that triggers a technology-adoption tipping point for air-conditioners. Conversely, increased frequency of extreme events might also lead to costly tipping points; for example, crossing a threshold in the frequency of crop failures might trigger a migration or conflict tipping point. Sufficiently large or frequent extreme events might also trigger an environmental policy tipping point that accelerates greenhouse gas mitigation, or that leads to the deployment of large-scale climate engineering technologies such as SRM [*Keith*, 2013; *Irvine et al.*, 2014].

Extreme events have the potential to serve as 'focusing events': sudden crises that result in calls for a remedy to reduce impacts of that crisis or chances of a future crisis. Natural disasters have long been seen as focusing events. Focusing events provide an opportunity for "interest groups, government leaders, policy entrepreneurs, the news media, or members of the public to identify new problems, or to pay greater attention to existing but dormant problems, potentially leading to a search for solutions in the wake of apparent policy failure" [*Birkland*, 1998]. As recognized by *Bentley* [2014], the acceleration of a trend towards a social tipping point often entails network feedbacks that amplify a message or behavior to many people [*Gladwell*, 2000]. The strength of the network matters: *Birkland* [1996] found that the expert community of scientists and government agencies around earthquakes is active and connected enough to keep earthquake safety on the national agenda between focusing events, whereas hurricanes have a sparse national expert network that results in hurricanes falling off the national agenda between events. In thinking more broadly about shifting U.S. public opinion and action on climate change, *Nisbet and Kotcher* [2009] suggested that a network of different types of opinion leaders that could be prepared before a climate-related focusing event and mobilized after might be effective in pushing public opinion towards a tipping point on climate action. However, to date, the negative feedback provided by groups leveraging public opinion leaders and networks to maintain policy stasis has generally been more effective [*Jasny et al.*, 2015; *Farrell*, 2016].

The study of climatically influenced social tipping elements is still in an early phase, but may play an important role in the integrated assessment of the cost of climate change. Environmental policy and technological tipping points that affect mitigation or climate engineering can influence the climatic trajectory that causes damage, while those that affect adaptation can influence the system resilience that modulates the translation of physical impacts into human costs. Migration and conflict tipping points, as we discuss below, may be important drivers of economic shocks. We suggest that it is important for the research community to more broadly survey the landscape of potential climate-related social tipping elements, to investigate networks and other social mechanisms driving relevant positive feedbacks, and to further develop statistical and systems-dynamic models characterizing these mechanisms. Statistical and dynamical models of social tipping elements have the potential to bring greater realism to the socio-economic projections of IAMs and to help move the representation of decision processes within IAMs away from the dictatorship of the infinitely lived representative agent.

## 5 Economic shocks with potential climate linkages

Tipping point research entered the IAM realm through an explicit though rather tenuous link to economic catastrophes in the DICE-99 model of *Nordhaus and Boyer* [*Nordhaus and Boyer*, 2000]. The previous two sections discussed climatic tipping elements and climatically sensitive social tipping elements, and suggested ways they might be incorporated into risk assessment. A complementary approach might be to start with the end-point of a large economic shock and work backwards, asking: What sort of phenomena do we know to cause large economic shocks? Through what pathways might climate change affect the probability of these phenomena? Some of these pathways involve tipping elements; not all do. We begin with the observation that the specific pathway, in how climate damages affect the economy, matters. We then cover several potential climate-economy shocks. For some, including capital-destroying meteorological disasters, civil wars, and temperature-linked hits to economic growth, climatic links are well-established. For some others, like financial crises, international wars on a nation's own soil, and large-scale political and economic restructuring, links cannot be excluded but are much less direct.

Economic damages can be effected through several different channels, including damages to output (as typically represented in IAMs), to capital stock or savings [*Fankhauser and Tol*, 2005], to labor productivity [*Graff Zivin and Neidell*, 2014; *Houser et al.*, 2015], to ecosystem services and other natural capital [*Sterner and Persson*, 2008], and to total factor productivity [*Moyer et al.*, 2013]. Some damages affect primarily the level of these factors, which is the standard assumption in most IAMs; others affect growth rates and could lead to vastly larger long-run damages [*Dell et al.*, 2012; *Heal and Park*, 2013; *Burke et al.*, 2015; *Moore and Diaz*, 2015]. Like gradual economic damages, economic shocks could be effected through any of these channels. For a full accounting of the economic impacts of climate damages, it is crucial to specify the particular channel. Although *Nordhaus* [1994] defined an economic catastrophe with respect to global economic output, world GDP has risen nearly continuously since at least 1950, and world GDP per capita has experienced only a few individual years of stagnation (1975, 1982, 1991, and 2009) [*The Maddison Project*, 2013; *Bolt and Zanden*, 2014]. To collect data on economic shocks, we must therefore narrow our scope to a national level (Table 2).

The large economic shocks with the most direct climate ties are associated with **capital-destroying meteorological disasters**. *Hsiang and Jina* [2014] showed that tropical cyclones cause a long-lasting (> 20 year) reduction in GDP. Among the countries ever hit by tropical cyclones, a "1-in-100 country-year" causes a persistent ~15% output reduction. Although *Hsiang and Jina* [2014] measured cyclones solely by their wind speed, a significant fraction of the damage caused by cyclones is flood-related, so it is reasonable to expect that sea-level rise will lead to more cyclone damage. Moreover, some studies indicate a regional or global increase in the number and intensity of tropical cyclones with climate change, which would make large shocks more frequent [*Emanuel*, 2013; *Knutson et al.*, 2013].

Though less directly influenced by climate than cyclones, **civil wars** both cause large economic shocks and also have an environmental connection [*Hsiang et al.*, 2011, 2013]. *Cerra and Saxena* [2008]'s analysis of 190 countries for the period 1960 to 2001 found that civil wars caused a ~6 ± 1% output loss after one year, with ~3 ± 3% persisting for at least a decade. Civil wars that coincided with a strengthening of executive power led to more severe economic crises: an output reduction of ~15 ± 6% that persisted for over ten years after the crisis started. Moreover, as previously mentioned, econometric results indicate that certain climatic conditions make civic conflict more likely. A meta-analysis of 31 studies indicated that a 1-standard deviation increase toward higher temperatures or more extreme rainfall leads to a ~14% increase in the frequency of civil conflict [*Hsiang and Burke*, 2013; *Hsiang et al.*, 2013]. Though this conclusion is not universally accepted, and though much work needs to be done to understand the mechanisms of the climate-conflict link [e.g., *Hsiang and Meng*, 2014; *Buhaug*, 2015], civil conflicts are a key area for future work assessing potential climate-economic shocks.

Warmer temperatures themselves also have the potential to create a slowly burning climate-economic shock through **temperature-induced effects on economic growth rates**. Empirical work shows that, in mid- and low-latitude countries (with an average annual temperature above ~13°C), higher annual average temperatures reduce economic growth rates [*Dell et al.*, 2009, 2012; *Burke et al.*, 2015]. Under high-end emissions projections (RCP 8.5) and a slow baseline growth scenario, *Burke et al.* [2015]'s mean estimated growth effect could make GDP per capita in nearly half of countries lower in 2100 than in 2010 – a genuine economic catastrophe for those countries, though not one tied to a specific triggering event. One possible pathway for this reduction in economic growth is the effect of extreme heat days on labor productivity [*Heal and Park*, 2013; *Graff Zivin and Neidell*, 2014]. Another contributing pathway is the well-established, nonlinear link between extreme heat and agricultural productivity [*Schlenker and Roberts*, 2008, 2009; *Roberts and Schlenker*, 2011; *Feng et al.*, 2012; *Tack et al.*, 2015].

Other historical triggers of large economic shocks have less clear potential for environmental triggers. *Cerra and Saxena* [2008] found that, after those associated with the combination of civil war and stronger executive power, the next largest economic shocks came from **twin currency/banking crises** (~10 ± 4% reduction in GDP ten years after the crisis starts). Currency crises occur when countries lack sufficient reserves to service debt or pay for critical imports, and are often resolved either through major devaluations, foreign-exchange market interventions, or austerity; banking crises are characterized by government-led closures, mergers, or takeovers of financial institutions, often in response to bank runs; and the most severe losses happen when the two occur together, as in the 1997 Asian financial crisis or in many regions in 2008 [*Kaminsky and Reinhart*, 1999].

Inspection of historical data shows that **international war on a nation's own soil** – thankfully, perhaps too rare for robust statistical analysis – carries a stark penalty as well. GDP per capita in France declined 50% from 1939 to 1944, and did not return to its 1939 level until 1948, while GDP per capita in Germany declined 60% from 1944 to 1946, recovering in 1955. GDP per capita in Iraq peaked in 1979 and fell to 38% of its peak value by 1989, after the end of the Iran-Iraq War; subsequent years of invasion and civil war has led it to a GDP per capita that in 2010 was 26% of its peak. **Large-scale political and economic restructuring** also has the potential to suppress GDP per capita: in the nascent USSR, GDP per capita declined 50% between 1917 and 1922, taking until 1926 to recover; in the ruins of the USSR, it declined 45% from 1989 to 1996, taking until 2006 to recover [*The Maddison Project*, 2013; *Bolt and Zanden*, 2014].

Although it may not be feasible to quantitatively assess the ways in which climate change could enhance the likelihood of financial crises, international wars, or political and economic revolutions, hypothetical 'worst-case' scenarios might highlight areas in need of quantitative investigation. Might simultaneous environmental shocks in different parts of the world, made more likely as a result of either a changing distribution of extremes with gradual climate change or as a result of a change in a tipping element, threaten the global financial system? Could such simultaneous shocks fuel war and conflict? Could climate change induce migration flows large enough to trigger institutional collapse in a major world power or make international war more likely? If qualitative analysis suggest these scenarios are plausible, then it would highlight the need for quantitative work on the correlation of extreme events and on migration.

In addition, worst-case scenarios involving both quantifiable and unquantifiable catastrophes might usefully supplement quantitative analysis for risk management purposes. While *Martin and Pindyck* [2015] discussed the importance of taking seriously tradeoffs among multiple 'catastrophes,' the potential for large economic shocks might necessitate entirely different decision criteria in addition to standard benefit-cost frameworks [*Garner et al.*, in press; *Hall et al.*, 2012; *McInerney et al.*, 2012; *Heal and Millner*, 2014; *Singh et al.*, 2015; *Wagner and Weitzman*, 2015]. Staying in the standard benefit-cost framework draws attention to the all-important element of discounting. If realizing a committed change takes many centuries or longer, discounting far-out damages becomes perhaps the most important aspect of evaluating the seriousness of any one particular tipping point [*Gollier*, 2012; *Arrow et al.*, 2013, 2014; *Giglio et al.*, 2015].

## 6 Conclusion: Pathways toward integrated assessment

Climatic tipping elements, climatically sensitive social tipping elements, and climate-economic shocks may all be important contributors to the costs of climate change; indeed, it is possible that they may be the largest contributors. Their incorporation into climate change risk analyses is therefore a crucial task that requires pursuing multiple research pathways.

One set of pathways starts with candidate climatic tipping elements and evaluates the climate-economic shocks that they might cause. While a number of studies have attempted to assess shocks associated with stylized tipping elements, more realistic approaches are possible. In particular, the growing body of empirical, econometric analyses of the effects of climate variability [e.g., *Dell et al.*, 2014], as well as an emerging set of sectoral models of impact processes [e.g., *Warszawski et al.*, 2014], have started to influence the assessment of more gradual climate change impacts. As discussed in section 3.4, experiments with empirical and process-based impact models, applied to scenarios in which the critical thresholds of different natural-system tipping elements are crossed, can help assess the economic hazards posed by state shifts and prioritize economic research on tipping elements that are quantitatively important. As noted therein, it is also important not just to consider tipping elements independently, but also to examine how tipping-element state shifts may influence both one another and also climate-economic shocks triggered by other causes. These experiments may show that some tipping elements have impacts that are small relative to more gradual climatic changes, and thus not a priority for integrated assessment research. They can also reveal how the economic influence of different tipping elements differs between regions and on different timescales.

Another set of pathways, discussed in section 4, expands the nascent body of work on the effects of climate change upon social tipping elements. Empirical evidence regarding policy thresholds, technology adoption, and technology costs can inform estimates of the costs of, efficacy of, and barriers to adaptation and mitigation, as well as of barriers to climate engineering. These estimates will improve assessments of the probability of different future scenarios of both climate change and social resilience to climate impacts. Empirical data on the relationships among climate change, migration and civil conflict may reveal costs that are larger than those associated many climatic tipping elements. A more comprehensive survey of potential climatically sensitive social tipping elements, with a focus on the role of positive feedbacks and the development of statistical and system-dynamic models, will help yield projections of human responses to climate change that are more realistic than those produced by models based upon a small number of idealized, rational actors.

A third set of pathways, discussed in section 5 and not entirely independent of the study of social tipping elements, starts with the concern that climate change may cause large-scale economic shocks and works backwards to identify mechanisms by which climate change may influence known shock triggers. Triggers with relatively clear mechanisms include meteorological disasters, civil conflict, and temperature effects on growth rates, but less obvious triggers like financial crisis and international wars should not be neglected. Both statistical and system-dynamic approaches may play important roles in tracing these mechanisms.

Fourth, in many cases, empirical and modeling studies may be able to identify potential state shifts or climate-economic shocks without being able to assess their likelihood, timing, or magnitude. Under such circumstances, structured expert elicitation, in which the probability estimates of experts are combined based upon calibrated assessments of the experts' reliability [*Kriegler et al.*, 2009; *Bamber and Aspinall*, 2013; *Oppenheimer et al.*, 2016], can temporarily fill in these gaps. But because conventional benefit-cost analysis performs poorly when expert disagreement on probability is too large, alternative approaches for robust decision making may be better suited to the context of tipping elements and climate-economic shocks [*Heal and Millner*, 2014]. One priority for policy research should be therefore focus on whether and how robust decision-making frameworks should displace benefit-cost analysis within existing, real-world decision-making processes. Advancing understanding of tipping elements and climate-economic shocks will require greater coordination among disciplines. By their nature, many critical thresholds involve pushing natural and/or social systems beyond the limits of behaviors to which we are accustomed, and thus require deep process insight informed by a diversity of empirical observations over a wide range of time scales. But understanding tipping points may be key to accurately assessing the societal and economic challenges posed by climate change, because they may be – as *Nordhaus and Boyer* [2000] suggest – one of the primary sources of climate change risk.

## Acknowledgments

This paper was supported by a gift to the Environmental Defense Fund from the Linden Trust for Conservation. It was originally produced as the discussion paper for the Climate Damages and Tipping Points Forum organized by Environmental Defense Fund and Resources for the Future, funded by the Litterman Family Foundation, and hosted at Arizona State University on 22 January 2016. The data on phrase frequency shown in Figure 1 were generated using the Google Books Ngram Viewer (https://books.google.com/ngrams) and Thomson Reuters Web of Science (https://webofknowledge.com). The model in Figure 2 is described in the Supporting Information. We thank the conference attendees for helpful discussion, and S. Bathiany, M. Brown, M. Buchanan, R. Cooke, H. Hausermann, P. Irvine, D. Keith, K. Keller, R. Litterman, A. McCright, P. McElwee, B. McKay, M. Oppenheimer, R. Pierrehumbert, D. Rasmussen, and D. Schrag for helpful comments and discussions.

**Table 1.** Illustrative candidate tipping elements

| Candidate Climatic Tipping Element | Main impact pathways | Potentially Gladwellian | Key Reference |
|---|---|---|---|
| *Atmosphere/ocean circulation* | | | |
| Atlantic Meridional Overturning Circulation | regional temperature, precipitation; global mean temperature; regional sea level | yes | [*Rahmstorf*, 2005] |
| Atmospheric superrotation | climate sensitivity (cloudiness) | yes | [*Caballero and Huber*, 2013] |
| El Niño-Southern Oscillation | regional temperature, precipitation | yes | [*Latif et al.*, 2015] |
| Regional North Atlantic convection | regional temperature, precipitation | yes | [*Drijfhout et al.*, 2015] |
| West African Monsoon | regional temperature, precipitation | yes | [*Dong and Sutton*, 2015] |
| *Cryosphere* | | | |
| Antarctic ice sheet | sea level; albedo | no | [*Schoof*, 2007] |
| Arctic sea ice | regional temperature, precipitation; albedo | yes | [*Li et al.*, 2013] |
| Greenland ice sheet | sea level; albedo | no | [*Robinson et al.*, 2012] |
| *Carbon cycle* | | | |
| Methane hydrates | greenhouse gas emissions | no | [*Archer et al.*, 2009] |
| Permafrost carbon | greenhouse gas emissions | no | [*Schuur et al.*, 2015] |
| *Ecosystem* | | | |
| Amazon rainforest | ecosystem services; greenhouse gas emissions | no | [*Jones et al.*, 2009] |
| Boreal forest | ecosystem services; greenhouse gas emissions; albedo | no | [*Jones et al.*, 2009] |
| Coral reefs | ecosystem services | yes | [*Hoegh-Guldberg et al.*, 2007] |
| *Socio-economic* | | | |
| Conflict/development | economic welfare | yes | [*Hsiang et al.*, 2013] |
| Environmental policy | adaptation/vulnerability; greenhouse gas emissions | yes | [*Baumgartner et al.*, 2014] |
| Migration | economic welfare | yes | [*Massey and Zenteno*, 1999] |
| Technology | adaptation/vulnerability; greenhouse gas emissions | yes | [*Rogers*, 2003] |

**Table 2.** Illustrative large economic shocks

| Economic catastrophe | Illustrative effect | Example Reference |
|---|---|---|
| Environmental disaster | ~15% output reduction for >20 years due to 1-in-100 country-year cyclone | [*Hsiang and Jina*, 2014] |
| Civil war | ~15% output reduction for >10 years if combined with strengthened executive power | [*Cerra and Saxena*, 2008] |
| Temperature-induced growth rate effects | Potential stalling of growth in warm countries with low productivity growth | [*Burke et al.*, 2015] |
| Twin currency/banking crises | ~10% output reduction for >10 years | [*Cerra and Saxena*, 2008] |
| International war on country's own soil | Transient per-capita output drop of > 50% in Europe during World War II | [*The Maddison Project*, 2013] |
| Large-scale political and economic restructuring | 45% drop in GDP/capita in Russia from 1989 to 1996 | [*The Maddison Project*, 2013] |

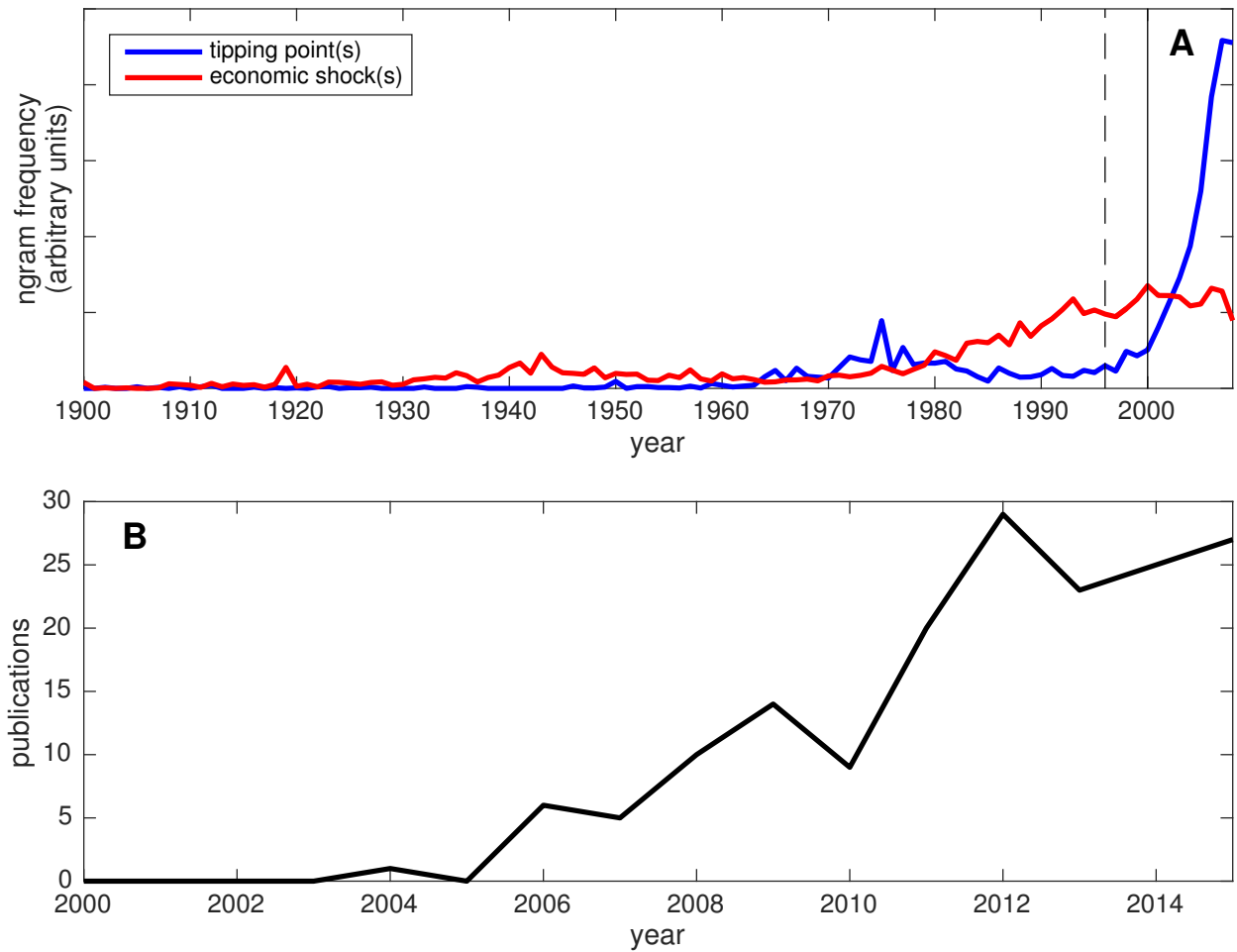

**Figure 1**. (a) Annual relative frequency (1900-2008) of the n-grams 'tipping point' (or 'tipping points') and 'economic shock' (or 'shocks') in the Google Books English language corpus, version 2 [*Michel et al.*, 2011]. The vertical dashed and solid black lines mark the publication of *Gladwell* [1996] and *Gladwell* [2000]. (b) Number of publications through the end of 2015 with the phrases 'tipping point' and 'climate change' in Web of Science (accessed 24 April 2016).

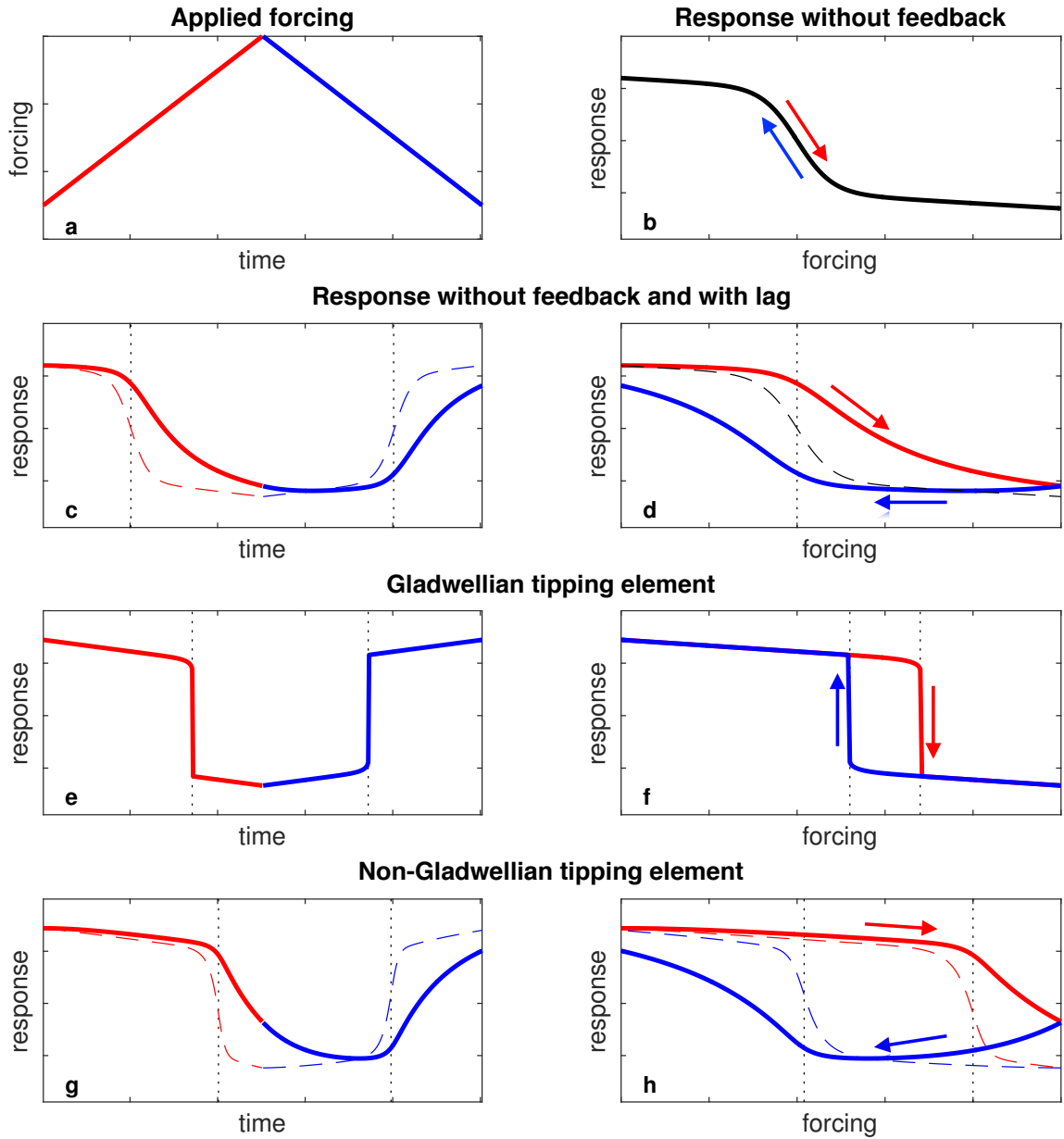

**Figure 2.** Illustration of the response of different types of tipping elements and non-tipping elements with thresholds to forcing. (a) Forcing applied over time in illustrative examples. In all plots, the red portion corresponds to the rising forcing and the blue portion to the falling forcing. In (b-h), the dotted vertical lines mark the time or forcing of the maximum rate of committed change; for tipping elements, this maximum rate defines the the critical threshold or 'tipping point.' (b) Response to forcing of a system with a threshold but no positive feedbacks. The same relationship between forcing and direct response occurs in all systems shown. A system that exhibits the response shown in (b) is not a tipping element. (c-d) Response without feedback, as in (b), with the addition of a lag, plotted as a function of (c) time and (d) forcing. This system is also not a tipping element. (e-f) Response of a Gladwellian tipping element, plotted as a function of (e) time and (f) forcing. (g-h) Response of a non-Gladwellian tipping element, plotted as a function of (g) time and (h) forcing. In (c-d) and (g-h), solid lines represent realized change, and dotted lines represent committed change.